\documentstyle[aps,prd,preprint]{revtex}

\def\half{{\textstyle{1\over2}}}

\def\quar{{\textstyle{1\over4}}}
\def\tfrac#1/#2{{\textstyle\frac#1#2}}

\def\mn{_{\mu\nu}}
\def\m{_\mu}
\def\n{_\nu}
\def\MN{^{\mu\nu}}
\def\d{\mathop{\rm d\null}\nolimits}
\def\pp{\ \ .}
\def\gt{$g=2$}
\def\eps{\varepsilon}

\def\beq{\begin{equation}}
\def\eeq#1{\label{#1}\end{equation}}

\newcommand{\LL}{{\cal L}}

\tighten

\begin{document}

\title{g = 2 as a Gauge Condition}

\footnotetext[1] {\baselineskip=16pt This work is supported in part by funds
provided by  the U.S.~Department of Energy (D.O.E.) under contracts
\#DE-FC02-94ER40818). \smallskip

\noindent
\mbox{}\hfill MIT-CTP-2664 \qquad  August 1997 
\hfill\break }

\author{R. Jackiw\footnotemark[1]\\
 \bigskip}

\address{Center for Theoretical Physics\\ 
Massachusetts Institute of Technology\\ Cambridge, MA
~02139--4307}

\maketitle
\begin{abstract}%
Charged matter spin-1 fields enjoy a nonelectromagnetic gauge
symmetry when interacting with vacuum electromagnetism,
provided their gyromagnetic ratio is 2.\vspace{.5in}

\centerline{To be submitted to {\it Physical Review D}.}
\end{abstract}

\setcounter{page}{0}
\thispagestyle{empty}

\newpage

It is agreed that the natural value for the gyromagnetic ratio~$g$ of an elementary
charged particle coupling to the electromagnetic field $F\mn$ is $g=2$ (in the
absence of small radiative corrections), so that the
Bargmann-Michel-Telegdi\cite{BMT} equation of motion for the spin vector
$S\m$ takes its simplest form
$$
\frac{\d S\m}{\d\tau} = \frac em F\mn S^\nu\pp
$$
Further reasons for this choice can be given:
\begin{itemize}
\item[(1)] Within particle physics theories, Weinberg\cite{W} has shown that the
\gt\ value must hold in tree approximation in order that scattering
amplitudes possess good high-energy behavior. Furthermore, Ferrara, Porrati, and
Telegdi\cite{FPT} have implemented this requirement of good high-energy
behavior in a Lagrangian framework, and regained \gt. 

\item[(2)] Spin-1/2 charged leptons carry \gt, and this confirms the view that
they are ``elementary" particles. The only known higher-spin, elementary charged
particle -- the W~boson -- possesses a gyromagnetic ratio
consistent with \gt. This is of course in agreement with the ``standard model" where
electromagnetism is fitted into a non-Abelian gauge group, which involves  
nonminimal electromagnetic coupling that results in \gt. \cite{ref:4}
\end{itemize}

In this Comment there is offered yet another reason for preferring \gt\ for
charged vector mesons: The kinetic term in a manifestly Lorentz-invariant
Lagrange density for spin-1 fields possesses a (nonelectromagnetic) gauge
invariance that removes redundant field degrees of freedom, which do not propagate
in the vacuum. This gauge invariance is preserved when the fields couple to
external, 
\emph{vacuum} electromagnetism ($\partial^\mu F_{\mu\nu}=0$) provided
\gt.

This simple observation is easily demonstrated. Consider the Lagrange density for
free complex vector fields $W_\mu$
\beq
 \LL_0 = -\half \left|\partial\m W\n -\partial\n W\m\right|^2\pp
\eeq{eq1}
[In the following, we consider massless fields, or alternatively in the case of
massive fields the discussion concerns the kinetic (derivative) portion of the
Lagrange density.]  Clearly $\LL_0$ possess the nonelectromagnetic gauge
invariance
\beq
W\m \to W\m + \partial\m \xi
\eeq{eq2}
where $\xi$ is complex. $W\m$ is coupled to electromagnetism by replacing
derivatives with covariant ones
\beq
\partial\m W\n \to D\m W\n\equiv(\partial\m+ieA\m)W\n
\eeq{eq3}
and allowing a further nonminimal interaction
\beq
 \LL = -\half {G^*}\MN G\mn + ie(g-1)F\MN W^*\m W\n
\eeq{eq4}
\beq
 G\mn\equiv D\m W\n - D\n W\m \pp
\eeq{eq5}
The nonminimal interaction ensures that charged vector particles carry
gyromagnetic ratio~$g$. When the electromagnetic field is source-free and
\gt, one verifies that $\LL$ remains invariant (up to total derivative terms)
against the nonelectromagnetic gauge transformation~(\ref{eq2}), provided the
derivatives are replaced by covariant ones,
 \beq
W\m \to W\m + D\m \xi\pp
\eeq{eq6}
As is well known, for massive fields with \gt\ the transversality condition 
$D^\mu W\m=0$ follows from the Euler-Lagrange field equation, while in the
massless case that condition can still be imposed thanks to the nonelectromagnetic
gauge symmetry ~(\ref{eq6}).

A similar situation holds for interactions with non-Abelian gauge potentials. When
the vector meson fields form (in general) a complex multiplet $W\m^i$, which
transforms under non-Abelian gauge transformations with the unitary
representation matrices~$U^{ij}$
\beq
W\m^i \to (U^{-1})^{ij} W\m^j
\eeq{eq7}
whose anti-Hermitian generators are $T_a$, $[T_a,T_b]=f_{abc}T_c$, then the
gauge covariant derivative is
\beq
 (D\m W\n)^i = \partial\m W\n^i + A\m^{ij}W\n^j
\eeq{eq8}
where the gauge potential is an element of the Lie algebra in this representation
\beq
A\m^{ij} =A\m^a T_a^{ij} \pp
\eeq{eq9}
The Lagrange density for these fields reads (apart from a possible mass term) 
\beq
\LL = -\half (G\mn^* )^i (G\MN)^i + (g-1)(W^{\mu *})^i  F\mn^{ij}(W^\nu)^j \pp
\eeq{eq10}
A nonminimal coupling to the gauge field strength is present, and one verifies that
the transformation
\beq
 W\m  \to W\m + D\m \Theta
\eeq{eq11}
changes $\LL$ only by total derivative terms when \gt\ and the gauge fields are
sourceless ($ D^\mu F\mn = 0$).

Additionally, let us note that in three-dimensional space-time and with vector
fields in the adjoint representation, there exists another term that is invariant
(apart from a total derivative) against~(\ref{eq11}). The relevant Lagrange
density reads
\beq
\LL = -\quar (G\mn)^a (G\MN)^a + f_{abc}(W^\mu)^a (W^\nu)^b
F\mn^c +m\eps^{\mu\alpha\beta} W\m^a F_{\alpha\beta}^a\pp
\eeq{eq12}
 Please observe that thanks to the Bianchi identity satisfied by $F_{\alpha\beta}:
D\m \eps^{\mu\alpha\beta} F_{\alpha\beta}=0$, the last term possesses the
symmetry~(\ref{eq11}) for arbitrary field strengths, not only sourceless ones. The
strength parameter $m$ carries dimensions of mass and the term has been posited
previously in a gauge- and parity-invariant mass generation mechanism for a
gauge theory, where also the gauge transformation~(\ref{eq11}) was
introduced \cite{JP}. 

The spin-2 case does not exhibit exactly the same behavior as above, yet
something similar, but less direct, does hold. Without interactions, but with a
mass term, the spin-2 equation of motion for a symmetric, second-rank
tensor
$h\mn$ can be taken as
\begin{eqnarray}
m^2(h\mn + \half g\mn h) &=& -\Box^2 h\mn + \partial\m h\n + \partial\n h\m
-\partial\m \partial\n h \nonumber\\
h\n \equiv \partial^\mu h\mn, && h\equiv h^\mu\m
 \label{eq13}
\end{eqnarray}
and the right side (massless part) enjoys the nonelectromagnetic gauge invariance
\beq
 h\mn \to h\mn + \partial\m \theta\n +\partial\n \theta\m \pp
\eeq{eq14}
Electromagnetic interactions can be included by promoting all derivatives to
covariant ones, ordered in a specific fashion, and adding a nonminimal
interaction. Thus one has 
\begin{eqnarray}
m^2(h\mn + \half g\mn h) &=& - D^2 h\mn + D\m h\n +  D\n h\m \nonumber\\
&& \quad {}-\half (D\m D\n + D\n D\m)h - ieg (F_{\mu\alpha} h\n^\alpha +
F_{\nu\alpha} h\m^a)
 \label{eq15}
\end{eqnarray}
 where now $h\n = D^\mu h\mn$.  With the ordering chosen above, the
gyromagnetic ratio is~$g$. Upon calculating the response of the right side to the
gauge covariant version of the substitution~(\ref{eq14})
\beq
 h\mn \to h\mn + D\m \theta\n + D\n \theta\m
\eeq{eq16}
one finds, with constant $F\mn$ ($\partial_\alpha F\mn=0$)
\beq
\Delta\mn = ie(2-g)(F\mn D^\alpha\theta\n + F_{\nu\alpha}D^\alpha\theta\m)
-ie(1+g)(F\mn D\n\theta^\alpha + F_{\nu\alpha}D\m\theta^\alpha)
\eeq{eq17}
so that for \emph{no} value~$g$ is invariance regained. $\bigl[$Note that for the
``minimal" value\cite{B} $g=\half$,
$$ 
\Delta\mn = ie\tfrac3/2 \Bigl(
        F_{\mu\alpha} (D^\alpha\theta\n -
        D\n\theta^\alpha) + F_{\nu\alpha} (D^\alpha\theta\m -
        D\m\theta_\alpha)
    \Bigr)
$$ 
the nonvanishing response involves the antisymmetric combination
$D_\alpha \theta_\beta - D_\beta \theta_\alpha$, while the definition of the
transformation in~(\ref{eq16}) makes use of the symmetric combination $D_\alpha
\theta_\beta + D_\beta \theta_\alpha$.$\bigr]$ 
However, one may improve the situation by the following (rather artificial)
consideration. Note that~(\ref{eq16}) implies that $h\m$ transforms as
\beq 
 	h\n \to D^2\theta\n + D^\mu D\n \theta\m
\eeq{eq18}
that is, the ordering of the noncommuting covariant derivatives is inherited from
previous definitions. But if we view $h\m$  as an independent quantity, we
can prescribe an arbitrary ordering in~(\ref{eq18}), which is equivalent to
modifying~(\ref{eq18}) by a multiple of $F_{\nu\mu}\theta^\mu$
\begin{eqnarray}
h\n &\to & D^2\theta\n + D^\mu D\n \theta\m + iec F_{\nu\mu}\theta^\mu
\nonumber\\ 
&& = D^2\theta\n + (1-c)  D^\mu D\n \theta\m + cD\n D^\mu\theta\m \pp
 \label{eq19}
\end{eqnarray}
Then the change in the kinetic part of the equation of motion~(\ref{eq15})
becomes
\beq
\Delta\mn = ie(2-g)(F_{\mu\alpha} D^\alpha\theta\n +
F_{\nu\alpha}D^\alpha\theta\m) -ie(1+g-c)(F_{\mu\alpha} D\n\theta^\alpha +
F_{\nu\alpha}D\m\theta^\alpha)
\eeq{eq20}
so that for the unique choice \gt\ the nonelectromagnetic gauge invariance is
maintained in the presence of constant fields, as long as the ordering
on~(\ref{eq19}) is taken with $c=3$. It remains an open question whether a
more fundamental/natural reason can be found for this \emph{ad hoc} ordering
prescription.

\nopagebreak

\end{document}